# Intelligent Radome Design Using Multilayer Metamaterial Structures to Realize Energy Isolation and Asymmetric Propagation of Electromagnetic Wave

Jing Yuan, Xiangkun Kong, *Member, IEEE*, Qi Wang, Chen Wu

*Abstract*—An intelligent radome utilizing composite metamaterial structures is presented and investigated in this article, which can realize energy isolation and asymmetric propagation of electromagnetic (EM) wave self-adaptively by controlling states of PIN diodes. The whole structure mainly consists of a broadband polarization-sensitive polarization converter (PC) and an active frequency selective rasorber (AFSR) switching between a transmission mode and absorption mode which is used as an energy-selective surface (ESS). Among them, the function of the PC is to make the EM waves transmit asymmetrically, and the purpose of AFSR is to make the high-power waves be reflected or absorbed, which depends on the polarization type of the wave. Thus, the radome can realize both asymmetric propagations of EM wave and electromagnetic shielding. The equivalent circuit models (ECM) and parametric studies are considered to explain the physical operating mechanism of PC and AFSR. The fabricated structure with 7×7 unit cells is experimentally demonstrated and the measured results agree with simulated results well. Considering the distinctive characteristic of self-actuation, the presented concept has the potential application in electromagnetic stealth and HPEMWs shielding to protect communication devices.

## I. INTRODUCTION

The phenomenon of asymmetric EM wave transmission is useful in designing the nonreciprocal EM wave devices, such as antenna radome, isolators and circulators[1]. But these devices are usually built by heavy and bulky material.

During the past decades, metamaterial has drawn much attention because of its unique properties that the natural material does not possess[2]-[4]. As an important application of metamaterial, chiral structure or metal wire grids structure can manipulate the polarization state of EM wave and transmit the EM wave asymmetrically. In paper[5],the twisted split-ring resonator patterns on both sides of the slab are used to transmit the linear polarized EM wave asymmetrically via polarization conversion. But this structure only can work at a resonant frequency. Then, a double I-shape structure is designed to achieve dual-band asymmetric transmission of linear polarization to overcome the above disadvantage[6]. With the rapid development of modern communication, broad bandwidth is necessary. The broadband or ultra wide band asymmetric propagation of EM wave is realized through chiral structure or metal mesh strips structure[7]-[10].

And recently, due to the rapid development of high-power microwave and electromagnetic pulse weapons, many modern electronic devices and communication systems may face threats[11]-[13].The most common way to protect information equipment is putting them in the shielding room. But this method could limit the functions of devices that need to have direct access to the outside. In order to solve this problem and realize self-actuated protection, Liu et al. presented an energy-selective surface (ESS) [14]-[17] using frequency selective surface (FSS) [18][19]. The FSS embedded PIN diodes is able to isolate HPEMWs while it can also keep the transmission performance at the working band. When the small signals are incident on the structure, the diodes array is at cutoff and the ESS is transparent to the signals; on the contrary, the ESS reflects or absorbs HPEMWs directly when the diodes are triggered by high power wave. Up to now, multifunctional AFSR and active FSS are designed[20]-[23] which can be applied in shielding HPEMWs. But these structures cannot propagate the wave asymmetrically. Moreover, Feng et al. proposed a one-way absorber for linearly polarized EM wave utilizing composite structures[24]. By combining PC and bi-directional polarization-sensitive absorber, the design realizes asymmetric EM wave transmission. But the whole

This work was supported by the Fundamental Research Funds for the Central Universities (No. kfjj20190406), the Fundamental Research Funds for the Central Universities (No. kfjj20180401), National Natural Science Foundation of China (61471368), Aeronautical Science Foundation of China (20161852016) and by Open Research Program in China's State Key Laboratory of Millimeter Waves (Grant No. K202027).

J. Yuan, X. Kong, Q. Wang and C. Wu are with the Key Laboratory of Radar Imaging and Microwave Photonics, Nanjing University of Aeronautics and Astronautics, Nanjing 210016, China (e-mail:1309157000@qq.com; xkkong@nuaa.edu.cn; 1040396539@qq.com; wuchen199701@163.com).
X. Kong is also with the State Key Laboratory of Millimeter Waves，Southeast University, Nanjing 210096, China

structure cannot protect the communication devices when HPEMWs attack it.

In this article, an intelligent radome, which can not only realize asymmetric propagation but also isolate HPEMWs, is presented. The radome consists of two parts: a PC and an AFSR. The broadband PC is constructed from three functional layers and two air layers, which can convert y-polarized incident waves to x-polarized transmitted waves and reflect the x-polarized incident waves directly when the wave comes from -z-direction. And, the situation is the opposite when the wave is from +z-direction. The polarization-sensitive AFSR is made up of slot arrays and also contains three layers. The AFSR switches between transmission mode and absorption mode by controlling the states of PIN diodes. By combining PC and AFSR, the intelligent radome can transmit the wave asymmetrically from 3.48GHz to 3.58GHz where insert loss (IL) is lower than 1dB when small signals are incident on the structure or the antenna emits signals. While HPEMWs impinge on the radome, the high power triggers diodes and HPEMWs are absorbed between 1.8GHz and 4.4GHz, realizing energy isolation. In order to simulate the incident HPEMWs and reduce extra negative effects, a parallel biasing configuration is designed to connect to the slots. Then equivalent circuit models(ECM) for the two parts are proposed to explain physical mechanisms. Finally, the presented structure was fabricated through the printed circuit board (PCB) technique and was measured in free space, exhibiting good agreement with simulations.

## II. BASIC STRUCTURE DESIGN

Fig.1 shows the working principles of the presented intelligent radome. Fig.1(a) depicts that when the common communication signals are incident on the structure or the antenna emits signals that are not powerful enough to trigger the threshold of PIN diodes, the structure realizes asymmetrical EM wave propagation at transmission band owning to sensibility to polarization. However, when EM wave containing high power impinges on the structure from outside, the proposed structure is opaque to HPEMWs that could be reflected or absorbed fully for different modes of polarization as Fig.1(b) exhibited, realizing energy isolation.

In Fig.2, the geometries of unit cell design and construction are given. Structure I is a wideband polarization-sensitive PC, as shown in Fig.2(a). It consists of three functional layers and two air layers. The first layer and the third functional layer consist of metal strips orthogonal to each other, which forms an Fabry-Pérot cavity to achieve broad bandwidth. The second functional layer containing an elliptic metal rotating 45° can generate resonance. When the y-polarized wave is incident on the PC from -z-direction, it can be converted into x-polarized wave; however, it reflects x-polarized incident wave directly. And the situation is opposite if the wave comes from +z-direction. The Fig.2(b) illustrates the structure of AFSR, which is acted as an ESS in this proposed design. The whole AFSR structure containing three structural layers and two air layers is constructed from slot arrays embedded on lumped elements and diode arrays. The first structural layer and the third structural layer are the same lossy layers which are spatial symmetric about the second layer in the aim to realize the same transmission characteristic of EM wave when the wave comes from two directions. The second layer is the lossless layer which serves as a passband FSS or metal ground depending on the state of PIN diodes.

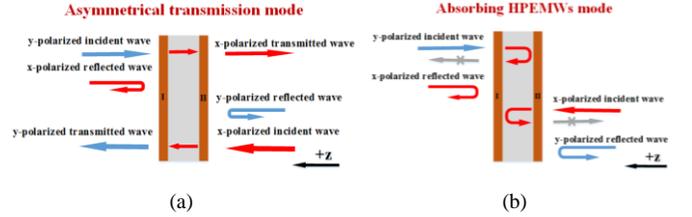

Fig.1 Working modes of the proposed structure in different states. (a) PIN diodes at OFF state. (b) PIN diodes at ON state.

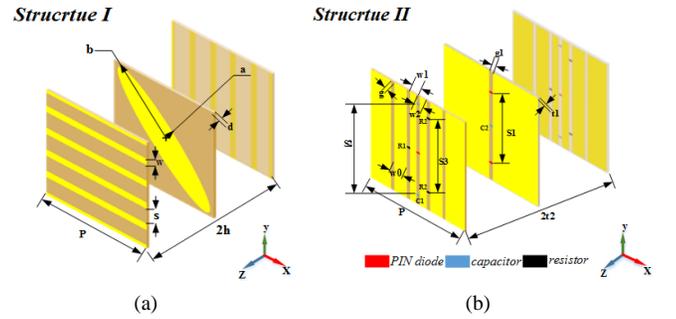

Fig.2 Unit cell geometries for the proposed structure. (a) broadband PC. (b) ESS.

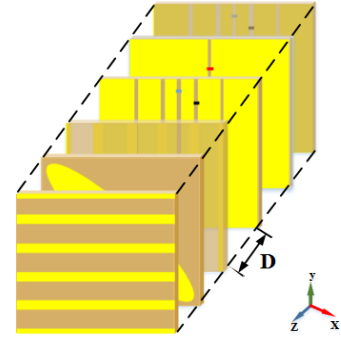

Fig.3 The whole structure of the intelligent radome.

As shown in Fig.3, based on these structures, the multifunctional radome can realize asymmetric EM wave transmission and energy isolation simultaneously. The final optimized geometric dimensions of the presented structure are as follows: $P$=38mm, $s$=5.25mm, $w$=2.25mm, $a$=5.25mm, $b$=24mm, $d$=2mm, $h$=12.75mm, $w0$=5mm, $w1$=2.5mm, $w2$=2.8mm, $g$=2mm, $g1$=1.6mm, $s1$=22.8mm, $s2$=36.2mm, $s3$=24.7mm, $t1$=0.5mm, $t2$=20mm, $D$=20.22mm, $R1$=80Ω, $R2$=30Ω, $C1$=1.4×$10^{-12}$F, $C2$=5×$10^{-9}$F. The F4Bs ($\varepsilon$=2.65, tan $\delta$=0.001) are chosen as the dielectric substrates. The PIN diode (SMP1345-079LF from SKYWORKS) is considered as a capacitance ($C_{off}$=0.15pF) in the OFF state and a small resistor

($R_{on}$=1Ω) in series with an inductor ($L_{on}$=0.7nH) in ON state in full-wave simulation.

### III. SIMULATION AND EQUIVALENT CIRCUIT MODEL

To fully understand the conversion mechanism of the wideband PC, a four ports network ECM is set up according to the basic principle of [25]. In Fig.4(a) and Fig.4(b), the port 1 and port 4 represent incident y-polarized wave and incident x-polarized wave, respectively. And the port 2 and port 3 represent transmitted co-polarized wave and transmitted cross-polarized wave for incident wave. The initial dimensions of the structure according to the circuit parameters can be obtained referring to [26][27] as follow formulas:

$$L_{eq} = -\mu_0 \frac{L}{2\pi} \log_2[\sin(\frac{\pi w}{2L})] \quad (1)$$

$$C_{eq} = -\varepsilon_0 \varepsilon_{eff} \frac{2L}{\pi} \log_2[\sin(\frac{\pi g}{2L})] \quad (2)$$

The $\mu_0$ is the permeability and $\varepsilon_0$ and $\varepsilon_{eff}$ represent permittivity and effective permittivity. $L$, $w$ and $g$ are the length, width and gap distance of the metal strips or patches. $Z_0$ and $Z_{sub}$ are the characteristic impedance of free space and spacer, respectively. Then by tuning and optimizing relevant electrical parameters of ECM slightly, it can be seen from compared simulation results in Fig.4(c) that the PC can convert y-polarized wave to x-polarized wave from 1.9GHz to 6.1GHz along -z-direction and conversion rate is up to about 100%. Still, the situation is opposite when the wave comes from +z-direction. Due to the polarization selection of metal grids structures, the PC can realize asymmetric transmission of EM wave.

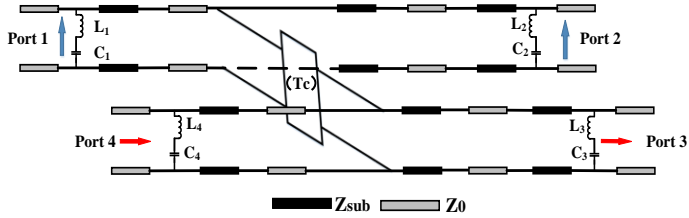

(a)

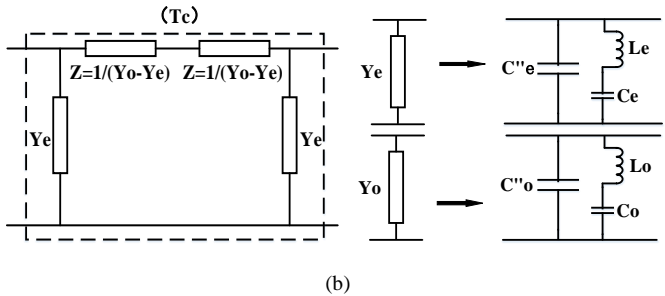

(b)

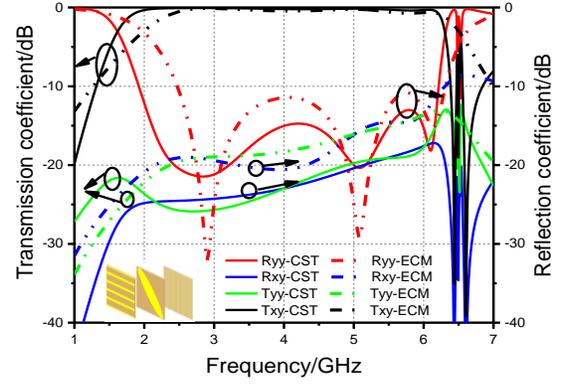

(c)

Fig.4 (a) The ECM of PC: $L_1=L_2$=1.9nH, $C_1=C_2$=0.569pF, $L_3=L_4$=165nH, $C_3=C_4$=5pF. (b) Pi-representation of the connection quadripole: C"e=0.053pF, Le=0.18nH,Ce=0.049pF,C"o=0.196pF,Lo=4.464nH,Co=0.13pF.(c)Simulation results of PC.

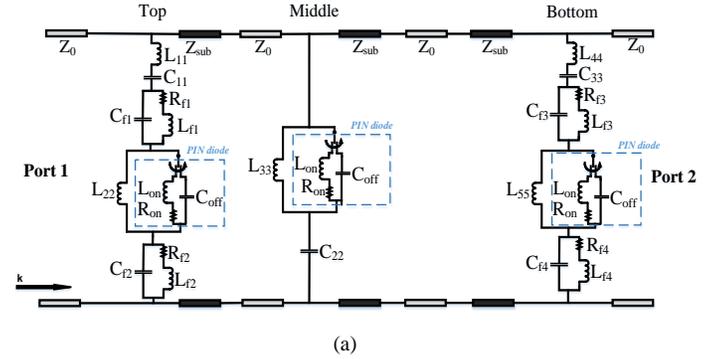

(a)

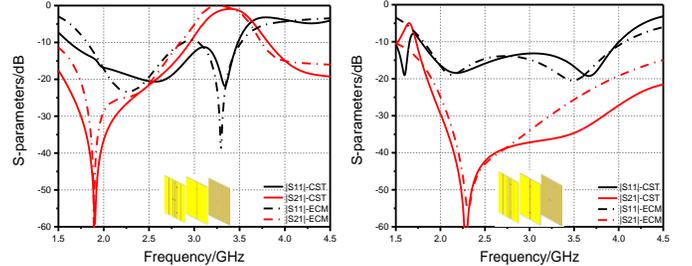

(b) (c)

Fig.5 The ECM of AFSR and the simulation results (a) ECM of structure:$L_{11}$= 6nH,$C_{11}$=0.17pF,$C_{f1}$=0.5pF,$R_{f1}$=48Ω,$L_{f1}$=6.95nH,$L_{22}$=4nH,$C_{off}$=0.6pF,$R_{ON}$=2Ω , $L_{ON}$=5nH,$C_{f2}$=0.2pF,$R_{f2}$=110Ω,$L_{f2}$=1.5nH,$L_{33}$=4.16nH,$C_{33}$=0.45pF,$L_{44}$=17.5nH $C_{f3}$=0.51pF,$R_{f3}$=52Ω,$L_{f3}$=22.9nH,$L_{55}$=3.52nH,$C_{f4}$=0.21pF,$R_{f4}$=110Ω,$L_{f4}$=1.5nH. (b) PIN diodes at OFF state (c) PIN diodes at ON state.

The ECM of the AFSR is shown in Fig.5(a), the top layer and bottom layer are the lossy layers and the middle layer is the lossless layer. The equivalent models of PIN diodes are also shown in the model. The resonant frequencies are caused by the slot arrays between the metal strips and are expressed as RLC circuits in the ECM. Fig.5(b) and (c) show the scattering parameters of the proposed AFSR from CST and ADS under normal incidence. The simulated results of CST agree well with that of ECM calculated by ADS. It can be shown that when the PIN diodes are at the cutoff, the structure acts as a FSR with a passband window behind a lower absorption band and the minimal insert loss (IL) of the passband window is only 0.6dB; when the PIN diodes are ON, the structure performs as an absorber which can absorb x-polarized wave between 1.8GHz and 4.4GHz covering S-band.

After optimizing the geometric dimensions and performances of PC and AFSR respectively, the combined structure is simulated in the CST. The distance between the two parts has an important effect on the resonance characteristics of structure because of mutual coupling[28].So designing a suitable distance is very important for the transmission band. The final results are plotted in Fig.6. When AFSR is combined with the broadband PC, the overall structure realizes switchable functionality between asymmetric propagation and shielding HPEMWs depending on the states of PIN diodes. The simulation results show that there is a passband with IL less than 1dB from 3.48GHz to 3.58GHz located behand the absorption band and the minimal IL is only 0.7dB at 3.54GHz when the structure is in a state of transmitting EM waves. However, when HPEMWs impinge on the surface, the PIN diodes are ON and HPEMWs are absorbed fully between 1.86GHz to 4.4GHz, realizing the function of isolating high energy .

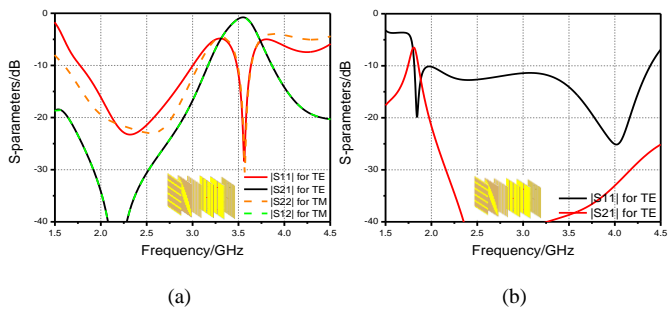

(a)          (b)

Fig.6 The simulation results of the whole structure (a) PIN diodes at OFF state (b) PIN diodes at ON state.

## IV. EXPERIMENTAL VERIFICATION

In order to demonstrate the performances of the proposed structure, a prototype was fabricated which consists of 7×7 unit cells with a size of 300 mm×300 mm and measured in an anechoic chamber as shown in Fig.7(a). The 1-20GHz standard horns connected to VNA(vector network analyzer) are used to transmit and receive EM waves. A voltage source provides bias voltage to PIN diodes in order to simulate the situation of HPEMWs. The 45mm-length nylon screws and foam spacers are employed to separate the layers and fix the whole structure. Furthermore, the absorbers blocks are placed around the prototype to avoid the edge diffraction.

The measured and simulated S-parameters are shown in Fig.7. The measured IL is 2.6dB and the resonant frequency shifts about 0.3GHz to higher frequency band when the PIN diodes are in OFF state. From the Fig.7 (b) and (c),it is clear that the structure can transmit EM wave asymmetrically when wave is from the two directions with different polarizations. On the other hand, when the PIN diodes are ON, the whole structure has more than 90% absorptivity from 2.3GHz to 4.4GHz. It means that it can shield the HPEMWs effectively. Some small differences between the simulations and measured results can be observed, which are caused by inaccurate sample fabrication, distortion of the sample pieces, and accuracy of the lumped elements. However, the lumped element set in the numerical simulation is equivalent to an ideal model according to the specification.

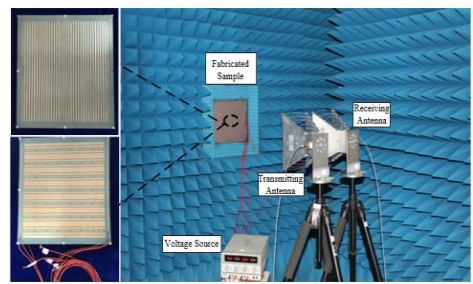

(a)

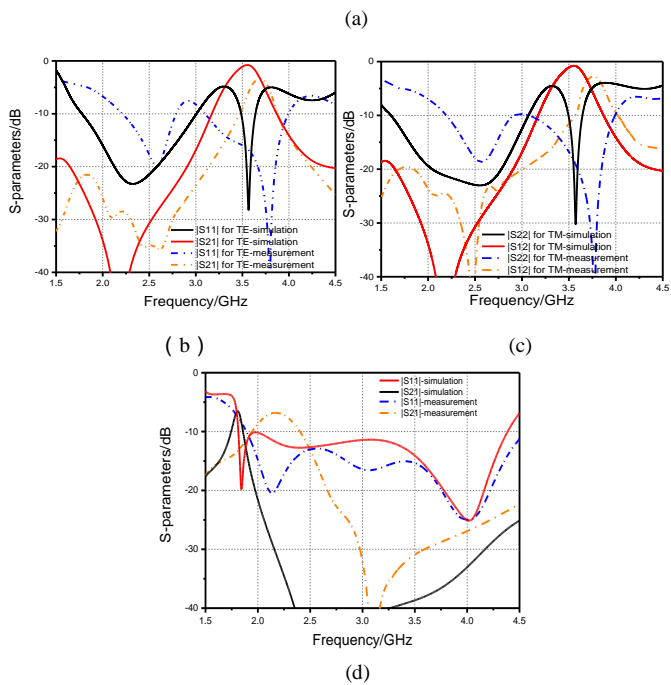

( b )          (c)

(d)

Fig.7 (a) The measurement setup of the structure (b) measured and simulated results from the -z direction for TE waves when PIN diodes are in OFF state (c) Measured and simulated results from the +z direction for TM waves when PIN diodes are in OFF state (d) Measured and simulated results when PIN diodes are in ON state.

## V.CONCLUSION

In this letter, we proposed an intelligent radome with multiple operating functions using the multilayer structures to realize asymmetric propagation of EM wave and shield the HPEWs at the same time depending on the states of PIN diodes. The broadband PC consisting of three functional layers and two air layers can transmit EM waves asymmetrically.

Then AFSR featured with lumped elements is made up of one lossless layer, two lossy layers and two air layers, which is performed as an ESS. The small signals which are not powerful enough to trigger PIN diodes can be transmitted asymmetrically. However, when HPEMWs impinge on the structure, the PIN diodes are ON and the HPEMWs are absorbed and shielded. The ECM is used to analyze their operating mechanism. Finally, the proposed intelligent radome is fabricated and measured. The measured results agree reasonably with the simulated results. The designed structure can be applied in modern communication systems and protect the antenna devices from high-power microwave and electromagnetic pulse weapons.